\documentclass[10pt,authoryear]{article}
\usepackage{appendix}
\usepackage{setspace}
\usepackage[font=small,  margin=2cm]{caption}
\usepackage[tbtags]{amsmath}
\usepackage{amsthm,amssymb,amsfonts}
\usepackage[numbers]{natbib}
\usepackage{multirow}
\usepackage{lscape}
\usepackage{subfigure}
\usepackage{makecell}
\usepackage{exscale}
\usepackage{booktabs}
\usepackage{array}
\usepackage{fullpage}
\usepackage{url}
\usepackage{algorithm}
\usepackage{algpseudocode}
\usepackage{bm}
\usepackage{smile}
\usepackage{mathtools}
\usepackage{wrapfig}
\usepackage{lipsum}
\usepackage{mathrsfs}
\usepackage{relsize}
\usepackage{dsfont}
\usepackage{multirow}
\usepackage{apalike}
\usepackage{chngcntr}
\usepackage[usenames,dvipsnames,svgnames,table]{xcolor}
\usepackage[colorlinks=true,
linkcolor=blue,
urlcolor=blue,
citecolor=blue]{hyperref}
\usepackage{ulem}
\newcommand{\li}{\uline{\hspace{0.5em}}}

\newcommand{\bargmin}{\mathop{\mathrm{arg\ min}}}
\numberwithin{equation}{section}
\numberwithin{theorem}{section}
\numberwithin{corollary}{section}
\counterwithout{asmp}{section}
\numberwithin{definition}{section}

\begin{document}

\title{\LARGE  Simultaneous Differential Network Analysis and Classification for High-dimensional Matrix-variate Data, with application to Brain Connectivity Alteration Detection and fMRI-guided Medical Diagnoses of Alzheimer's Disease}

	\author{Hao Chen\thanks{School of Statistics, Shandong University of Finance and Economics, Jinan, China; },~~Ying Guo\thanks{Department of Biostatistics and Bioinformatics, Rollins School of Public Health,
Emory University, Atlanta, USA; },~~Yong He\thanks{Institute for Financial Studies, Shandong University, Jinan, China; Email:{\tt heyong@sdu.edu.cn}.},~~Jiadong Ji\footnotemark[1],\\~~ Lei Liu\thanks{Division of Biostatistics, Washington University in St.Louis, St. Louis, USA;},~~Yufeng Shi\footnotemark [3], ~~Yikai Wang\footnotemark [2],~~ Long Yu\footnotemark[5],~~ Xinsheng Zhang\thanks{Department of Statistics, School of Management, Fudan University, Shanghai,  China; },\\~~ for the  Alzheimer's  Disease  Neuroimaging  Initiative\thanks{ Data  used  in  preparation  of  this  article  were  obtained  from  the  Alzheimer's  Disease  Neuroimaging  Initiative  (ADNI)  database  (adni.loni.usc.edu).  As  such,  the  investigators within the ADNI contributed to the design and implementation of ADNI and/or provided data but  did  not  participate  in  analysis  or  writing  of  this  report.  A  complete  listing  of  ADNI investigators can be found at:http://adni.loni.usc.edu/wp-content/uploads/how\li to\li apply/ADNI\li Acknowledgement\li List.pdf}}	
	\date{}	
	\maketitle

\textbf{Abstract:} Alzheimer's disease (AD) is the most common form of dementia, which causes problems with memory, thinking and behavior. Growing evidence has shown that the brain connectivity network experiences alterations for such a complex disease. Network comparison, also known as differential network analysis, is thus particularly powerful to reveal the disease pathologies and identify clinical biomarkers for medical diagnoses (classification). Data from neurophysiological measurements are multi-dimensional and in matrix-form, which poses major challenges in brain connectivity analysis and medical diagnoses. Naive vectorization method is not sufficient as it ignores the structural information within the matrix.  In the article, we adopt the Kronecker product covariance matrix framework to capture both spatial and temporal correlations of the matrix-variate data while the temporal covariance matrix is treated as a nuisance parameter. By recognizing that the strengths of network connections may vary across subjects, we develop an ensemble-learning procedure, which  identifies the differential interaction patterns of brain regions between the AD group and the control group and conducts medical diagnosis (classification) of AD  simultaneously. %We also provide a robust version of the two-stage method, which performs well in the presence of heavy-tailed data or outliers.
Thorough simulation study are conducted to assess the performance of the proposed method. We applied the proposed procedure to functional connectivity analysis of fMRI dataset related with Alzheimer's disease. The hub nodes and differential interaction patterns identified are consistent with existing experimental studies, and satisfactory out-of-sample classification performance is achieved for medical diagnosis of Alzheimer's disease.
A R package for implementation is available at \url{https://github.com/heyongstat/SDNCMV}.

\vspace{0.2em}


\section{Introduction}

\begin{figure}[H]
	\centerline{\includegraphics[width=16cm,height=8cm]{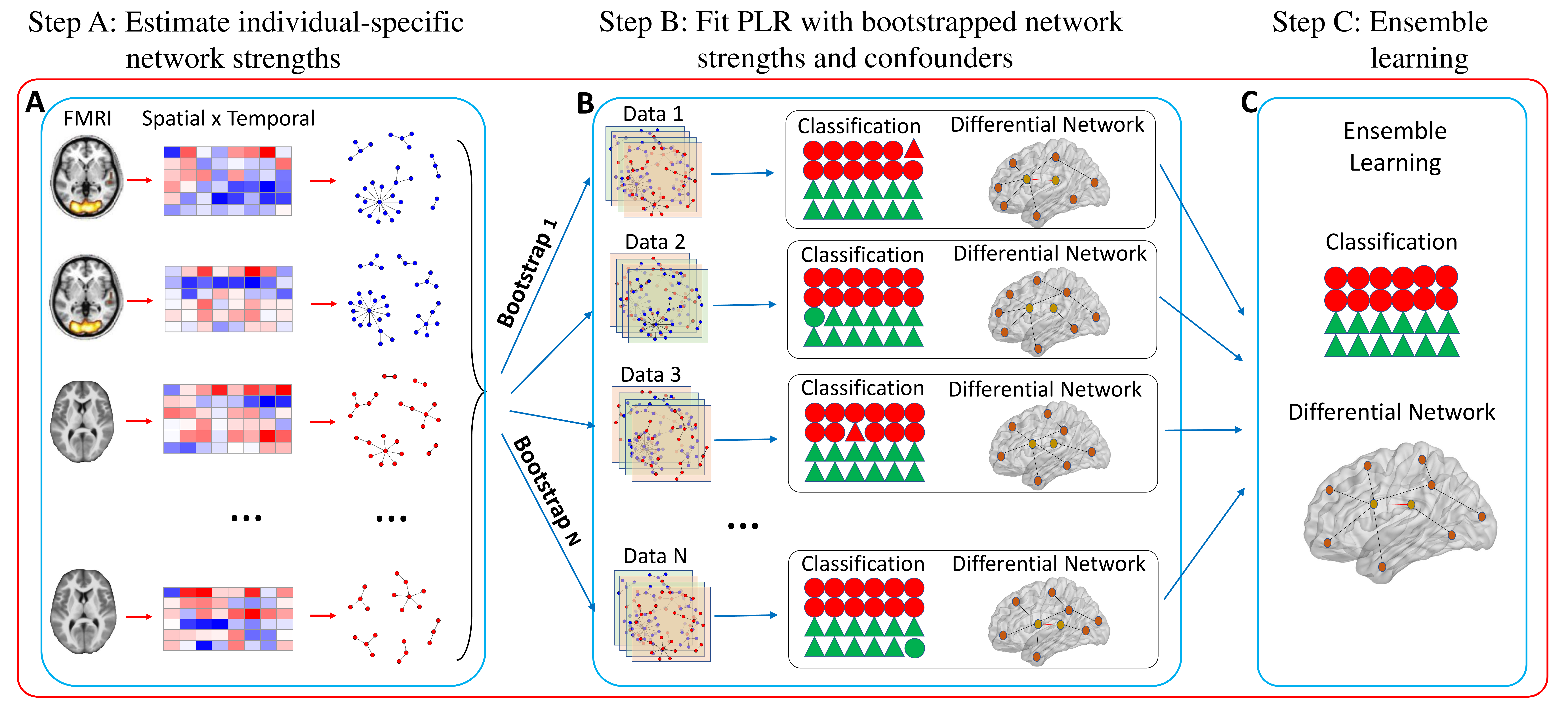}}
	\caption{Workflow of the SDNCMV method: (A) Estimate individual-specific network strengths with the Spatial$\times$ Temporal data; (B) Combine individual network information and confounders to fit Penalized Logistic Regression (PLR) with bootstrap samples (C) Ensemble learning with the bootstrap results from the PLRs.
	}
	\label{fig:workflow}
\end{figure}

Alzheimer's disease (AD) is  the most common neuro-degenerative disease. For AD diagnosis, neuroscience researchers often resort to Brain Connectivity Analysis (BCA) to reveal the underlying pathogenic mechanism through correlations in the neurophysiological measurement of brain activity. Functional Magnetic Resonance Imaging (fMRI), which records the blood oxygen level dependent (BOLD) time series, has been the mainstream imaging modalities to study brain functional connectivity.  The observed fMRI data have the spatial by temporal matrix structure, where the rows correspond to the brain regions while the columns correspond to the time points, see the fMRI and matrix data in part (A) of Figure \ref{fig:workflow}.

Growing evidence has shown that the brain connectivity network experiences alterations with the presence of Alzheimer's disease \citep{RyaliEstimation,Higgins2019a,Xia2018matrix}.  Differential network analysis or network comparison has been an important way to provide deep insights into disease pathologies, see \cite{Yuan2015Differential,tian2016identifying,ji2016powerful,yuan2016powerful,ji2017jdinac,Zhang2017Incorporating,He2018a,grimes2019integrating}.
Most literature focuses on differential network modelling for vector-form data. For the observed fMRI spatial by temporal matrix data, directly applying the existing differential network analysis methods for vector-form data would ignore the intrinsic matrix structure and may result in unreliable conclusion. \cite{Zhu2018multiple} employed non-convex penalization to tackle the estimation of multiple graphs from matrix-valued data, while \cite{Xia2018matrix} formulated the problem as testing the equality of individual entries of partial correlation matrices, both under a matrix normal distribution assumption.

An ensuing problem is the image-guided medical diagnosis of AD, i.e., to classify a new subject into the AD or control group from the fMRI data. Classification or discriminant analysis is a classical problem in statistics. There have been a good number of standard vector-valued predictor classification methods, such as logistic regression and linear discriminant analysis, and corresponding modified versions to cope with the high-dimensionality of predictors, see for example, \cite{bickel2004some,Zhu2004Classification,HuiRegularization,Young2008Penalized,fan2008high,cai2011direct,mai2012direct,Yong2016Discriminant,MaGlobal}.
For the fMRI matrix data, the naive idea, which first vectorizes the matrix and then adopts the standard vector-valued predictor classification, does not take advantage of the intrinsic matrix structure and thus leads to low classification accuracy. To address this issue, several authors presented logistic regression-based methods and linear discriminant criterion for classification with  matrix-valued predictors: \cite{Zhou2014Regularized} proposed a nuclear norm penalized likelihood estimator of the regression coefficient matrix in a generalized linear model; \cite{Li2D,ZhongMatrix,Molstad2019A}  modified Fisher's linear discriminant criterion for matrix-valued predictors.

In the article, we propose a method which achieves Simultaneous Differential Network analysis and Classification for Matrix-Variate data (SDNCMV). The SDNCMV is indeed an ensemble learning algorithm, which involves three main steps. Firstly, we propose an individual-specific spatial graphical model to construct the between-region network measures (connection strengths), see part (A) in Figure \ref{fig:workflow}. In practice, we measure the between-region network strengths by a specific transformation function of the partial correlations in order to separate their values between two groups well. In the second step, with the constructed individual between-region network strengths and confounding factors, we adopt the bootstrap technique and train the Penalized Logistic Regression (PLR) models with bootstrap samples, see part (B)  in Figure \ref{fig:workflow}. Finally, we ensemble the results from the bootstrap PLRs  to boost the classification accuracy and network comparison power,  see part (C)  in Figure \ref{fig:workflow}.

The advantages of  SDNCMV over the existing state-of-the-art methods lie in the following aspects. First, as far as we know, the SDNCMV is the first proposal in the neuroscience literature to achieve network comparison and classification for matrix fMRI data simultaneously. Second, the SDNCMV  performs much better than the existing competitors in terms of classification accuracy and network comparison power, especially when the two populations share similar mean (usually demeaned) while differ in population (partial) correlation structure - note that almost all the existing classification methods focus on the population mean difference with common correlation structure. Third, the SDNCMV is an ensemble machine learning procedure, which makes a strategic decision based on the different fitted PLRs with bootstrap samples. It is thus more robust and powerful. Fourth, the SDNCMV addresses an important issue in brain network comparison, i.e., it can adjust confounding factors such as age and gender, which has not taken into full account in the past literature. The results are illustrated both through simulation studies and the real fMRI data of  Alzheimer's disease.

The rest of the paper is organized as follows. In Section 2, we present the two-stage procedure which achieves classification and network comparison simultaneously. %In Section 3, we propose a robust version of the two-stage procedure.
In Section 3, we assess the performances of our method and some state-of-the-art competitors  via the Monte Carlo simulation studies. Section 4 illustrates the proposed method through an AD study. We summarize our method and present some future directions in Section 5.

\section{Methodology}

\subsection{Notations and Model Setup}
In this section, we introduce the notations and model setup.
We adopt the following  notations  throughout of the paper. For any vector $\ba=(a_1,\ldots,a_d)^\top \in \RR^d$, let $\|\ba\|_2=(\sum_{i=1}^d a_i^2)^{1/2}$, $\|\ba\|_1=\sum_{i=1}^d |a_i|$.  { Let $\Ab=[a_{ij}]$ be a square matrix of dimension $d$, define  $\|\Ab\|_0=\sum_{i=1}^d\sum_{j=1}^d I(a_{ij}\neq 0)$, $\|\Ab\|_1=\sum_{i=1}^d\sum_{j=1}^d |a_{ij}|$, $\|\Ab\|_{\infty}=\max|a_{ij}|$. We denote the trace of $\Ab$ as $\text{Tr}(\Ab)$ and let $\text{Vec}({\Ab})$ be the vector obtained by stacking the columns of $\Ab$. Let $\text{Vec}(b_{ij})_{j>i}$ be the operator that  stacks the columns of the upper triangular elements of
matrix $\Bb=(b_{ij})$ excluding the diagonal elements to a vector. For instance, $\Bb=(b_{ij})_{4\times 4}$, then $\text{Vec}(b_{ij})_{j>i}=(b_{12},b_{13},b_{23},b_{14},b_{24},b_{34})^\top$.
 The notation $\otimes$ represents Kronecker product. For a set $\mathcal{H}$, denote by $\#\{\mathcal{H}\}$ the cardinality of $\mathcal{H}$. For two real numbers $p,q$,  define $p\vee q=\max(p,q)$.}

 Let $\Xb_{p\times q}$ be the spatial-temporal matrix-variate data from fMRI with $p$ spatial locations and $q$ time points. We assume that the spatial-temporal matrix-variate $\Xb_{p\times q}$ follows the matrix-normal distribution with the Kronecker product covariance structure defined as follows.

\begin{definition}
 We say $\Xb_{p\times q}$ follows a matrix normal distribution with the Kronecker product covariance structure $\bSigma=\bSigma_{T}\otimes\bSigma_{S}$, denoted as
   {\setlength{\abovedisplayskip}{3pt}
   \setlength{\belowdisplayskip}{3pt}
   \[
   \Xb_{p\times q}\sim \cN_{p,q}(\Mb_{p\times q},\bSigma_{T}\otimes\bSigma_{S}),
   \]
   }
   if and only if $\text{Vec}(\Xb_{p\times q})$ follows a multivariate normal distribution with mean $\text{Vec}(\Mb_{p\times q})$ and covariance $\bSigma=\bSigma_{T}\otimes\bSigma_{S}$,
   where  $\bSigma_{S}\in \RR^{p\times p}$ and $\bSigma_{T}\in \RR^{q\times q}$ denote the covariance matrices of $p$ spatial locations and $q$ times points, respectively.
\end{definition}

   The matrix-normal distribution framework  is scientifically relevant in neuroimaging analysis and BCA study,  see, for example, \cite{Yin2012Model,Chenlei2012Sparse,Zhou2014Gemini,Xia2017Hypothesis,Zhu2018multiple,Xia2018matrix}. Under the matrix-normal distribution assumption of $\Xb_{p\times q}$, we have
   {\setlength{\abovedisplayskip}{3pt}
   	\setlength{\belowdisplayskip}{3pt}
   \[
   \text{Cov}^{-1}\big(\text{Vec}(\Xb_{p\times q})\big)=\bSigma_{T}^{-1}\otimes\bSigma_{S}^{-1}=\bOmega_{T}\otimes\bOmega_{S},
   \]}where $\bOmega_{S}\in\RR^{p\times p}$ and $\bOmega_{T}\in\RR^{q\times q}$ denote the spatial precision matrix and temporal precision matrix, respectively. $\bSigma_{S}$ and $\bSigma_{T}$  are only identifiable up to a scaled factor. In fact, in the brain connected network analysis, the partial correlations or equivalently the scaled precision matrix elements is a commonly adopted correlation measure \citep{Peng2009Partial,Zhu2018multiple}. In addition, the primary interest in BCA study is to infer the connectivity
network characterized by the spatial precision matrix $\bOmega_{S}$ while the temporal precision matrix $\bOmega_{T}$ is of little interest. Under the matrix normal framework, a region-by-region spatial partial correlation matrix, $\bR_{S}=\bD_{S}^{-1/2}\bOmega_{S}\bD_{S}^{-1/2}$, characterizes the  brain connectivity graph,  where $\bD_{S}$ is the  diagonal matrix of $\bOmega_{S}$. Brain connectivity analysis is in essence to infer the spatial partial correlation matrix $\bR_{S}$. In fact, the advantage of partial correlation has been recognized in the neuroscience community, as it measures the direct connectivity between two regions and  avoids spurious effects in network modeling, see  \cite{Smith2012The,WangAn2016}.

In the current article, we focus on the Brain Connectivity Alteration Detection (BCAD) of Alzheimer's disease, i.e., to identify the differences in the spatial partial correlation matrices of the AD group and the control group.  The most notable feature that differentiates our proposal from the existing BCAD-related literature is that {we take the variations of the strengths of spatial network connections across subjects into account}. At the same time, we also focus on the fMRI-guided medical diagnoses of Alzheimer's disease. Let $\Xb_{\gamma}\in\RR^{p\times q},\gamma=1,\ldots,n_1$ be the subjects from the AD group and $\Yb_{\gamma^\prime},\gamma^\prime=1,\ldots,n_2$ from the control group, which are all spatial-temporal matrices. We assume that $\Xb_{\gamma}\sim \cN_{p,q}(\Mb_{X}^{(\gamma)},\bSigma_{T_X}^{(\gamma)}\otimes\bSigma_{S_X}^{(\gamma)})$ and $\Yb_{\gamma^\prime}\sim \cN_{p,q}(\Mb_{Y}^{(\gamma^\prime)},\bSigma_{T_Y}^{(\gamma^\prime)}\otimes\bSigma_{S_Y}^{(\gamma^\prime)})$, which indicates the individual-specific between-region connectivity strengths.
\begin{remark}
Most existing literature assume that the individuals in the same group share the same spatial matrix while our proposal allows individual heterogeneity.  Recent medical studies provide evidence for  the presence of substantial heterogeneity of covariance network connections between individuals and subgroups of individuals, see for example, \cite{ChenAge2011,AlexanderImaging2013,Xie2020Identifying}.
\end{remark}

\subsection{The SDNCMV Procedure}

In this section, we present the detailed procedure for Simultaneous Differential Network analysis and Classification for Matrix-Variate data (SDNCMV). The SDNCM involves the following three main steps: (1) construct the individual-specific between-region network measures; (2) train the
Penalized Logistic Regression (PLR) models with bootstrap samples; (3) ensemble results from the bootstrap PLRs to boost the classification
accuracy and network comparison power. We give the details in the following subsections.

\subsubsection{Individual-specific between-region network measures}
In the section, we introduce the procedure to construct the individual-specific between-region network measures. We focus on the AD group while the control group can be dealt similarly.
Recall that $\Xb_{\gamma}\in\RR^{p\times q}$ is  the $\gamma$-th subject from the AD group, and let $\bx^{\gamma}_{\cdot \nu}\in\RR^p$ be the  $\nu$-th column of $\Xb_\gamma$, respectively. The individual spatial sample covariance matrix is simply
\begin{equation}\label{equ:samplecov}
  \hat\bSigma_{S_X}^{(\gamma)}=\frac{1}{q-1}\sum_{\nu=1}^q(\bx_{\cdot \nu}^\gamma-\bar\bx_{\gamma})(\bx_{\cdot \nu}^\gamma-\bar\bx_{\gamma})^\top,
\end{equation}
where $\bar\bx_{\gamma}=1/q\sum_{\nu=1}^q\bx_{\cdot \nu}^\gamma$. We assume that $p$ is larger than or comparable to $q$, which is typically the case for fMRI data.  We also assume that the $\bOmega_{S_X}^{(\gamma)}=(\bSigma_{S_X}^{(\gamma)})^{-1}$ is a $p\times p$ sparse matrix and estimate it by the Constrained
$\ell_1$-minimization for Inverse Matrix Estimation (CLIME) method by \cite{cai2011constrained}. That is,

\begin{equation}\label{equ:clime}
\tilde\bOmega_{S_X}^{(\gamma)}=\bargmin_{\bOmega}\|\bOmega\|_1, \ \ \text{subject to} \ \ \|\hat\bSigma_{S_X}^{(\gamma)}\bOmega-\Ib\|_\infty\leq \lambda,\ \ \bOmega\in\RR^{p\times p},
\end{equation}
where $\lambda$ is a tuning parameter. In the optimization problem in (\ref{equ:clime}), the symmetric condition is not imposed on $\bOmega$, and  the solution is not symmetric in general. We construct the final CLIME estimator $\hat\bOmega_{S_X}^{(\gamma)}=(\hat\Omega_{S_X,ij}^{(\gamma)})$  by symmetrizing $\tilde\bOmega_{S_X}^{(\gamma)}=(\tilde\Omega_{S_X,ij}^{(\gamma)})$ as follows. We compare the pair of non-diagonal entries at symmetric positions $\tilde\Omega_{S_X,ij}^{(\gamma)}$ and  $\tilde\Omega_{S_X,ji}^{(\gamma)}$, and assign the one with  smaller magnitude to both entries. That is,
   \[
   \hat\Omega_{S_X,ij}^{(\gamma)})=\hat\Omega_{S_X,ji}^{(\gamma)})
   =\tilde\Omega_{S_X,ij}^{(\gamma)}I(|\tilde\Omega_{S_X,ij}^{(\gamma)}|\leq|\tilde\Omega_{S_X,ji}^{(\gamma)}|)
   +\tilde\Omega_{S_X,ji}^{(\gamma)}I(|\tilde\Omega_{S_X,ij}^{(\gamma)}|>|\tilde\Omega_{S_X,ji}^{(\gamma)}|).
   \]
One may also use other individual-graph estimation methods such as Graphical Lasso \citep{friedman2008sparse}. We adopt CLIME for the sake of computational efficiency, which is very attractive for high-dimensional data as it can be easily implemented by linear programming. We select the tuning parameter $\lambda$ by the {\it Dens} criterion function proposed by \cite{WangAn2016}. We use the \textsf{R} package ``DensParcorr" to implement the procedure for inferring individual partial correlation matrix, which allows the user to specify the {desired density level $\varphi$}. We set the  the desired density level at $\varphi=0.5$ in practice.

For each subject in the AD group, we can obtain an estimator of the individual spatial precision matrix by the CLIME procedure, that is, $\hat\bOmega_{S_X}^{(\gamma)}, \gamma=1,\ldots,n_1$. Parallelly, we can obtain $\hat\bOmega_{S_Y}^{(\gamma^\prime)}, \gamma^\prime=1,\ldots,n_2$ for the control group subjects. We then scale the spatial precision matrices to obtain the partial correlation matrices
\[
 \hat\Rb_{S_X}^{(\gamma)}=(\hat\Db_{S_X}^{(\gamma)})^{-1/2}\hat\bOmega_{S_X}^{(\gamma)}(\hat\Db_{S_X}^{(\gamma)})^{-1/2}, \ \  \hat\Rb_{S_Y}^{(\gamma^\prime)}=(\hat\Db_{S_Y}^{(\gamma^\prime)})^{-1/2}\hat\bOmega_{S_Y}^{(\gamma^\prime)}(\hat\Db_{S_Y}^{(\gamma^\prime)})^{-1/2}, \ \ \gamma=1,\ldots, n_1, \ \  \gamma^\prime=1,\ldots, n_2,
\]
where $\hat\Db_{S_X}^{(\gamma)}$, $\hat\Db_{S_Y}^{(\gamma^\prime)}$ are the diagonal matrix of $\hat\bOmega_{S_X}^{(\gamma)}$, $\hat\bOmega_{S_Y}^{(\gamma^\prime)}$, respectively. We then define the individual-specific between-region network measures as follows:
\[
\hat W_{S_X,ij}^{(\gamma)}=\frac{1}{2}\log \left(\frac{1+\hat R_{S_X,ij}^{(\gamma)}}{1-\hat R_{S_X,ij}^{(\gamma)}}\right), \ \ \hat W_{S_Y,ij}^{(\gamma^\prime)}=\frac{1}{2}\log \left( \frac{1+\hat R_{S_Y,ij}^{(\gamma^\prime)}}{1-\hat R_{S_Y,ij}^{(\gamma\prime)}}\right), \ \ 1\leq i<j<p,
\]
where $\hat R_{S_X,ij}^{(\gamma)}$ and $\hat R_{S_Y,ij}^{(\gamma^\prime)}$ are the $(i,j)$-th element of $\hat\Rb_{S_X}^{(\gamma)}$, $\hat\Rb_{S_Y}^{(\gamma^\prime)}$, respectively.
In fact, the defined between-region network measures are the Fisher transformation of the estimated partial correlation. Fisher transformation is well known as an approximate variance-stabilizing transformation, which alleviates the possible effects of skewed distributions and/or outliers and contributes to the outstanding performance of the PLR in the second stage.
%In the case that $p\ll q$, we can simply omit the CLIME procedure and estimate the spatial precision matrices as the inverse of the sample covariance matrices.

\subsubsection{Penalized logistic regression}
Logistic regression is a classical statistical model  for a binary classification problem. In this section we introduce the penalized logistic regression model and the bootstrap procedure.

Let $n=n_1+n_2$ and let the binary response variable be denoted as $Z$ and its observations are $Z_1,\ldots,Z_n$, where
\[
Z_k=\left\{
\begin{array}{ccc}
1     &      & k\in \text{AD};\\
0    &      & k\notin \text{AD};
\end{array} \right. \ \  k=1,\ldots,n.
\]
Denote $P$ as the probability of the event $Z=1$, i.e., $P=\Pr(Z=1)$. The second-stage logistic model for Alzheimer's disease outcome is:

\begin{equation}\label{equ:logit}
\text{logit}(P)=\log\left(\frac{P}{1-P}\right)=\sum_{m=1}^M\eta_mQ_m+\sum_{i=1}^p\sum_{j>i}^p\beta_{ij}W_{S,ij},
\end{equation}
where $\bQ=(Q_1,\ldots,Q_m)^\top$ denote the confounder covariates (e.g. age and gender) and   $W_{S,ij}$ is the Fisher transformation of the spatial partial correlation  between the $i$-th and $j$-th regions $R_{S,ij}$, i.e., $$W_{S,ij}=\frac{1}{2}\log\left(\frac{1+R_{S,ij}}{1-R_{S,ij}}\right).$$

Note that it adds up to $M+p(p-1)/2$ variables in (\ref{equ:logit}), which could be very large when $p$ is large. Thus we are motivated to consider a sparse penalty on the coefficients $\{\beta_{ij},1\leq i<j\leq p\}$. Finally, the Penalized Logistic Regression (PLR) model for estimating $\bbeta=\text{Vec}(\beta_{ij})_{j>i}$ is as follows:

\begin{equation}\label{equ:penalized}
  (\hat{\bm{\eta}},\hat\bbeta)=\bargmin_{\bm{\eta},\bbeta}\left\{\frac{1}{n}\sum_{k=1}^n\bigg[-Z_k\big(\bm{\eta}^\top\bQ_k+\bbeta^\top \bW_k\big)+\log\Big(1+\exp\big(\bm{\eta}^\top\bQ_k+\bbeta^\top \bW_k\big)\Big)\bigg]+P_\lambda(\bbeta)\right\},
\end{equation}
where $\bm{\eta}=(\eta_1,\ldots,\eta_M)^\top$, $\bQ_k=(Q_{k1},\ldots,Q_{km})^\top$ is the $k$-th observation of $\bQ$, $P_{\lambda}(\cdot)$ is a penalty function and $\bW_k$ is the individual-specific network strengths estimated
from the first-stage model, i.e.,
\[
\bW_k=\left\{
\begin{array}{cccccccccccccccc}
\text{Vec}(\hat W_{S_X,ij}^{(k)})_{j>i}     &      & 1\leq k\leq n_1;\\
\text{Vec}(\hat W_{S_Y,ij}^{(k-n_1)})_{j>i}    &      & n_1+1\leq k\leq n.
\end{array} \right.
\]
 The penalty function $P_\lambda(\cdot)$ is selected as the elastic net penalty \citep{HuiRegularization} to balance  the $\ell_1$ and $\ell_2$ penalties of the lasso and ridge methods, i.e.,
\[
P_\lambda(\bbeta)=\lambda\Big(\alpha\|\bbeta\|_1+(1-\alpha)\|\bbeta\|_2^2\Big),
\]
where $\alpha\in[0,1]$ is a tuning parameter. The tuning parameters can be selected by the Cross-Validation (CV) in practice.

If the coefficient $\beta_{ij}\neq 0$, there exists an  edge  between the  brain regions $i$ and $j$ in the differential network, which can be used by PLR to discriminant the AD subjects from the control subjects. Thus, with the estimated support set of  $\hat\bbeta$, we can recover the differential edges and at the same time, the subjects in the test set are classified based on the fitted $\hat P$. That is how our proposal achieves Simultaneous Differential Network analysis and Classification for Matrix-Variate data (SDNCMV). It is worth pointing out that we can modify the model in (\ref{equ:logit}) such that it includes the original variables $\Xb_\gamma$ as a part of the confounders $\bQ$ and also add a penalty for $\bQ$. In the current paper, we mainly focus on the predictive ability of the individual-specific network strengths irrespective of the original variables.

\subsubsection{Ensemble Learning  with bootstrap}

In this section, we introduce a bootstrap-based ensemble learning method which further boosts  the classification accuracy and the network comparison power. With the $n_1$ subjects from the AD group and $n_2$ subjects from the control group, we randomly sample $n_1$ subjects from the AD group and  $n_2$ samples from the control group separately, both with replacement,  and then conduct the two steps as introduced in the last two sections. We repeat the re-sampling $B$ times and then we denote the regression coefficients as $\{\hat\bbeta^{(b)}=\text{Vec}(\hat\beta_{ij}^{(b)})_{j>i},b=1,\ldots,B\}$ and the out-sample classification label $\hat Z^{(b)}\in\{0,1\}$. We classify the new test sample as AD if $\hat P_B=1/B \sum_{b=1}^BI(\hat Z^{(b)}=1)>0.5$. Similarly, to boost the network comparison power, we calculate the differential edge weight for each pair of nodes $(i,j)$, defined as $\theta_{ij}=\sum_{b=1}^B I(\hat \beta_{ij}^{(b)}\neq 0), 1\leq i<j\leq  p$. {For a pre-specified threshold $\tau$,  we believe that there exists the edge $(i,j)$ in the differential network if $\theta_{ij}>\tau$. A  subjective way to determine $\tau$ is simply to take the value $B/2$. We also recommend to determine $\tau$ by a scree plot, see  the real example in Section 4.  } The simulation study in the following section shows that the bootstrap-assisted ensemble learning boosts the classification accuracy and the network comparison power considerably. The entire SDNCMV procedure is summarized in Algorithm \ref{alg:first}.

\begin{algorithm}[H]
	\caption{Ensemble Learning Algorithm for the SDNCMV}\label{alg:first}
	{\bf Input:}  $\cD_{Train}=\Big\{\Xb_\gamma,\gamma=1,\ldots,n_1,\Yb_{\gamma^\prime},\gamma^\prime=1,\ldots, n_2, \bQ_k, k=1,\ldots,n$\Big\},  $\cD_{Test}$\\
	{\bf Output:} Predicted labels of the test samples in $\cD_{Test}$,   differential edge weights $\theta_{ij},1\leq i<j\leq p$.
	\begin{algorithmic}[1]
  \Procedure{}{}
		\State Perform the CLIME procedure in (\ref{equ:clime}) for each subject in  $\cD_{Train}$ and $\cD_{Test}$.

		\State Calculate the network strengths $\bW_k$ for subjects in $\cD_{Train}$ and $\cD_{Test}$.
\For{ $b\leftarrow 1$ {\bf to} $B$}

        \State Sample $\{(\bQ_k,\bW_k),k=1,\ldots,n\}$ from $\cD_{Train}$ with replacement, and obtain bootstrap samples $\{(\bQ_k^{(b)},\bW_k^{(b)}),k=1,\ldots,n\}$.

        \State Solve the PLR in (\ref{equ:penalized}) with $\{(\bQ_k^{(b)},\bW_k^{(b)}),k=1,\ldots,n\}$, and obtain $(\bm{\hat\eta}^{(b)},\hat \bbeta^{(b)})$.

         \State Calculate the predicted labels for each subject in  $\cD_{Test}$ with $(\bm{\hat\eta}^{(b)},\hat \bbeta^{(b)})$, and  obtain $\hat Z^{(b)}$.
\EndFor

\State Classify the test subject to AD if $\hat P_B=1/B \sum_{b=1}^BI(\hat Z^{(b)}=1)>0.5$.

\State Let $\theta_{ij}=1/B\sum_{b=1}^B I(\hat \beta_{ij}^{(b)}\neq 0)$ and the differential network edges are estimated as $\{(i,j):\theta_{ij}>\tau, 1\leq i<j\leq  p\}$.
  \EndProcedure
	\end{algorithmic}
\end{algorithm}
\section{Simulation Study}\label{sec:simu}
In this section, we conducted simulation studies to investigate the  performance of the proposed method SDNCMV in terms of classification accuracy and network comparison power. In section \ref{sec:SS-1}, we introduce the synthetic data generating settings, and show the classification accuracy and network comparison results in Section \ref{sec:SS-2} and Section \ref{sec:SS-3}, respectively.

\subsection{Simulation Settings}\label{sec:SS-1}
In this section we introduce the data generating procedure for numerical study.  We mainly focus on the predictive ability of the  between-region network measures without regard to the confounder factors.
 To this end,  the data were generated from matrix normal distribution with  mean zero, i.e., for AD group, we generated $n_1$ independent samples $\Xb_{\gamma}$ $(\gamma=1,2,\dots,n_1)$ from $\cN_{p,q}\Big(\zero, \bSigma_X^{(\gamma)}\Big)$ with $\bSigma_X^{(\gamma)}=\bSigma_{T_X}^{(\gamma)}\otimes\bSigma_{S_X}^{(\gamma)}$; and for the control group, we generated $n_2$ independent samples $\Yb_{\gamma^\prime}$ $(\gamma^\prime=1,2,\dots,n_2)$ from $\cN_{p,q}\Big(\zero, \bSigma_Y^{(\gamma^\prime)}\Big)$ with $\bSigma_Y^{(\gamma^\prime)}=\bSigma_{T_Y}^{(\gamma^\prime)}\otimes\bSigma_{S_Y}^{(\gamma^\prime)}$ and the scenarios for the covariance matrices $\bSigma_X^{(\gamma)},\bSigma_Y^{(\gamma^\prime)}$ are introduced in detail below.

For the temporal covariance matrices: $\bSigma_{T_X}^{(\gamma)}=(\sigma_{{T_X},i,j}^{(\gamma)})_{q \times q}$ and $\bSigma_{T_Y}^{(\gamma^\prime)}=(\sigma_{{T_Y},i,j}^{(\gamma^\prime)})_{q \times q}$, the first structural type is the Auto-Regressive (AR) correlation, where $\sigma_{{T_X},i,j}^{(\gamma)} =0.4^{|i-j|}$ and $\sigma_{{T_Y},i,j}^{(\gamma^\prime)} =0.5^{|i-j|}$. The second structural type is Band Correlation (BC), where $\sigma_{{T_X},i,j}^{(\gamma)} =1/{(|i-j|+1)}$ for $|i-j|\le 4$ and 0 otherwise and $\sigma_{{T_Y},i,j}^{(\gamma^\prime)} =1/{(|i-j|+1)}$ for $|i-j| \le 6$ and 0 otherwise. The temporal covariance matrices of the individuals in the same group are exactly the same for simplicity.

\begin{figure}[]
	\centerline{\includegraphics[width=16cm,height=6 cm]{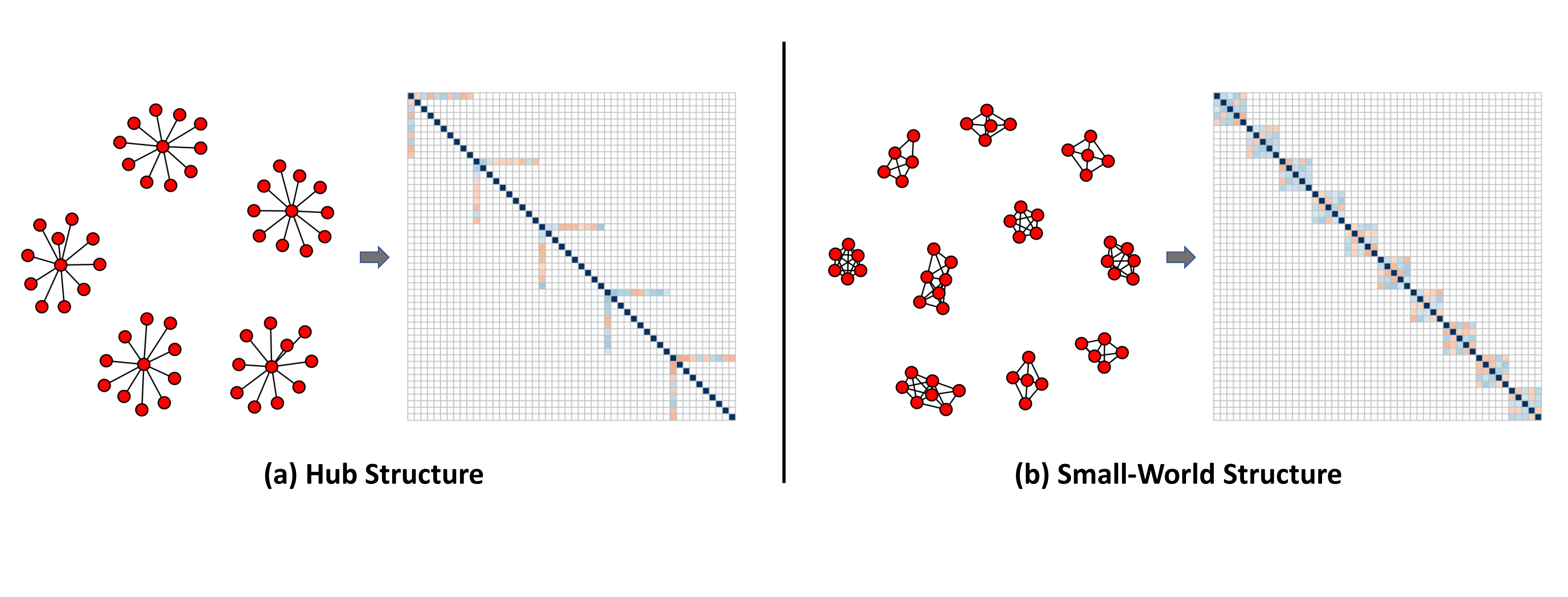}}
	\caption{Graph structures and the corresponding heat maps of correlation matrices considered in our simulation studies: (a) Hub structure; (b) Small-World structure}
	\label{fig:graphstructure}
\end{figure}

For the spatial covariance matrices, we first construct the  graph structure $\text{G}_{S}$. We consider two types of graph structure: the Hub structure and the Small-World structure.  We resort to \textsf{R} package ``huge" to generate Hub structure with 5 equally-sized and non-overlapping  graph and use \textsf{R} package ``rags2ridges" to generate Small-World structure with 10 small-world graph and 5\% probability of rewiring. The two graph structures and the corresponding heat maps of the partial correlation matrix are shown in Figure \ref{fig:graphstructure}. For further details of these two graph structures, one may refer to \cite{zhao2012huge} and \cite{Wieringen2016Ridge}.
Then, based on the graph structure $\text{G}_{S}$, we generate two base matrices,  $\bOmega_{S_X}$ for AD group and $\bOmega_{S_Y}$ for control group. In detail, we determine the positions of non-zero elements of matrices $\bOmega_{S_X}$ and $\bOmega_{S_Y}$ by $\text{G}_{S}$. Then we filled the non-zero positions in matrix $\bOmega_{S_X}$ with random numbers from a uniform distribution with support $[-2,-1] \cup [1,2]$. We randomly selected two blocks of $\bOmega_{S_X}$ and changed the sign of the element values to obtain $\bOmega_{S_Y}$. To ensure that these two matrices are positive definite, we let $\bOmega_{S_X}=\bOmega_{S_X}+(|\lambda_{\min}(\bOmega_{S_X})|+0.5)\Ib$ and $\bOmega_{S_Y}=\bOmega_{S_Y}+(|\lambda_{\min}(\bOmega_{S_Y})|+0.5)\Ib$. The differential network is in essence modelled as $\bDelta=\bOmega_{S_X}-\bOmega_{S_Y}$.
With the base matrices $\bOmega_{S_X},\bOmega_{S_Y}$, we then generate the individual-specific precision matrices. We generated $n_1$ independent matrices $\bPsi_{X}^{(\gamma)}$ by changing the non-zero elements in matrix $\bOmega_{S_X}$ to random values from the normal distribution $\cN(0, 0.02)$. Finally, we  get $n_1$ individual-specific precision  matrices by $\bOmega_{S_X}^{(\gamma)} = \bOmega_{S_X} + \bPsi_{X}^{(\gamma)}$,  with the same structure but different elements. The covariance matrices $\bSigma_{S_X}^{(\gamma)}$ are set to be $(\bOmega_{S_X}^{(\gamma)})^{-1}$ for $t=1,\ldots,n_1$. By a similar procedure, we get precision matrices $\bOmega_{S_Y}^{(\gamma^\prime)}$ and covariance matrices $\bSigma_{S_Y}^{(\gamma^\prime)}$ for $\gamma^\prime=1,\ldots,n_2$.

   To sum up, we considered  four covariance structure scenarios for $\bSigma_{X}^{(\gamma)},\gamma=1,\ldots,n_1$ and  $\bSigma_{Y}^{(\gamma^\prime)}, \gamma^\prime=1,\ldots,n_2$:
 \begin{itemize}
    \item \textbf{Scenario 1: } $\bSigma_{X}^{(\gamma)}=\bSigma_{T_X}^{(\gamma)} \otimes \bSigma_{S_X}^{(\gamma)}$ and $\bSigma_{Y}^{(\gamma^\prime)}=\bSigma_{T_Y}^{(\gamma^\prime)} \otimes \bSigma_{S_Y}^{(\gamma^\prime)}$, where $\bSigma_{T_X}^{(\gamma)}$ and $\bSigma_{T_Y}^{(\gamma^\prime)}$ are AR covariance structure type and $\bSigma_{S_X}^{(\gamma)}$ and $\bSigma_{S_Y}^{(\gamma^\prime)}$ are based on $\text{G}_{S}$ with Hub structure.

    \item \textbf{Scenario 2: } $\bSigma_{X}^{(\gamma)}=\bSigma_{T_X}^{(\gamma)} \otimes \bSigma_{S_X}^{(\gamma)}$ and $\bSigma_{Y}^{(\gamma^\prime)}=\bSigma_{T_Y}^{(\gamma^\prime)} \otimes \bSigma_{S_Y}^{(\gamma^\prime)}$, where $\bSigma_{T_X}^{(\gamma)}$ and $\bSigma_{T_Y}^{(\gamma^\prime)}$ are AR covariance structure type and $\bSigma_{S_X}^{(\gamma)}$ and $\bSigma_{S_Y}^{(\gamma^\prime)}$ are based on $\text{G}_{S}$ with Small-World structure.

    \item \textbf{Scenario 3: } $\bSigma_{X}^{(\gamma)}=\bSigma_{T_X}^{(\gamma)} \otimes \bSigma_{S_X}^{(\gamma)}$ and $\bSigma_{Y}^{(\gamma^\prime)}=\bSigma_{T_Y}^{(\gamma^\prime)} \otimes \bSigma_{S_Y}^{(\gamma^\prime)}$, where $\bSigma_{T_X}^{(\gamma)}$ and $\bSigma_{T_Y}^{(\gamma^\prime)}$ are BC covariance structure type and $\bSigma_{S_X}^{(\gamma)}$ and $\bSigma_{S_Y}^{(\gamma^\prime)}$  are based on $\text{G}_{S}$ with Hub structure.

    \item \textbf{Scenario 4: } $\bSigma_{X}^{(\gamma)}=\bSigma_{T_X}^{(\gamma)} \otimes \bSigma_{S_X}^{(\gamma)}$ and $\bSigma_{Y}^{(\gamma^\prime)}=\bSigma_{T_Y}^{(\gamma^\prime)} \otimes \bSigma_{S_Y}^{(\gamma^\prime)}$, where $\bSigma_{T_X}^{(\gamma)}$ and $\bSigma_{T_Y}^{(\gamma^\prime)}$ are BC covariance structure type and $\bSigma_{S_X}^{(\gamma)}$ and $\bSigma_{S_Y}^{(\gamma^\prime)}$ are based on $\text{G}_{S}$ with Small-World structure.

  \end{itemize}

We set $p \in \{100, 150\}$, $q \in \{50, 100\}$ and $n_1=n_2=30$, the Bootstrap times $B=200$ and all simulation results are based on 100 replications. For the assessment of classification accuracy, we also
generate $n_1^{test}=n_2^{test}=n_1=n_2=30$ test samples independently for each replication, where $n_1^{test},n_2^{test}$ are the sizes of test samples from the AD group and the control group respectively.
Also note that even $p=100$, there are $p(p-1)/2=4950$ variables in the second-stage penalized logistic regression.
 %When $p$ is large, we can first adopt the Sure Independent Screening (SIS)  procedure by \cite{Fan2008Sure}  and then fit the PLR using the variables in the active set (the survived variables).
%In the case that $p<q$, we omit the CLIME procedure and directly estimate the  individual spatial precision matrices by the inverse of the sample covariance matrices in the first-stage.

\subsection{Classification Accuracy Assessment}\label{sec:SS-2}
To evaluate the classification performance of our method, we considered four competitors including the  {Matrix-valued  Linear Discriminant Analysis (MLDA) by \cite{Molstad2019A}}   and three classical vector-valued machine learning methods: Random Forest (vRF), Support Vector Machine (vSVM) and vector-valued Penalized Logistic Regression (vPLR). To implement the vector-valued machine learning methods, we simply vectorize the  matrices $\Xb_{\gamma}, \Yb_{\gamma^\prime}$ of dimension $p\times q$ and treat them as observations of a random vector of dimension $pq$. For fair comparison, we also consider the Random Forest (RF),  Support Vector Machine (SVM) and Penalized Logistic Regression (PLR) methods with the same input variables as SDNCMV, i.e., the $p(p-1)/2$ variables $W_{S,ij},1\leq i<j\leq p$. It is worth noting that these methods have not been used for classification with $W_{S,ij}$ in neuroscience community. We adopted the existing R packages to implement these competitors, i.e.,  ``MatrixLDA" for MLDA, ``randomForest" for RF and vRF, ``e1071" for SVM and vSVM, and ``glmnet" for PLR and vPLR. We compute the misclassification error rate on the same test set of size $n_1^{test}+n_2^{test}=60$ for each replication to compare classification accuracy.

\begin{table}[h!]\caption{{Misclassification rates (standard errors) averaged of the 100 replications (\%) for Scenario 1-4} }\label{tab:Classification}
\centering
\scalebox{0.88}{
\renewcommand{\arraystretch}{1.2}
\begin{tabular*}{17.5cm}{ccrccccccc}
\toprule[2pt]

   $(p, q)$ && SDNCMV   & RF  &SVM & PLR  & vRF  &vSVM & vPLR  & MLDA\\
   \hline
   & & \multicolumn{8}{c}{\textbf{Scenario 1}}\\

   $(100, 50)$ &        & 0.0 (0.2)	 &0.8 (1.2)  &8.8 (4.9) &1.2 (1.5) &49.8 (6.2)  &49.4 (6.0) &49.7 (6.4) & 50.6 (6.1)\\
   $(150, 50)$ &        & 0.0 (0.2)	 &1.4 (1.6) &17.5 (6.8)  &1.3 (1.7) &50.3 (7.2) &50.4 (6.5)  &49.8 (6.9) &49.0 (8.4)\\
   $(100, 100)$ &      &0.0 (0.0)	 &0.0 (0.2) &3.3 (2.6)  &0.9 (1.8) &50.2 (6.2) &49.0 (6.9)  &50.6 (6.4) &50.2 (7.8)\\
   $(150, 100)$ &      &0.0 (0.0)	 &0.1 (0.4) &4.5 (3.3)  &1.0 (1.6) &50.2 (5.9) &48.8 (5.8)  &49.5 (6.7) &49.5 (7.9)\\

   & & \multicolumn{8}{c}{\textbf{Scenario 2}}\\

   $(100, 50)$ &        &0.0 (0.2)	 &0.1 (0.5) &2.7 (2.5)  &1.6 (2.2) &49.3 (6.4) &48.7 (5.8)  &51.0 (5.9) &50.3 (6.2)\\
   $(150, 50)$ &        &0.0 (0.2)   &0.1 (0.4) &3.3 (3.1)  &1.4 (1.8) &49.1 (7.4) &45.9 (5.2)  &49.6 (6.6) &49.9 (6.9)\\
   $(100, 100)$ &      &0.0 (0.0)	 &0.0 (0.0) &0.4 (0.8)  &0.8 (1.5) &50.5 (5.4) &49.8 (5.8)  &50.3 (6.8) &50.7 (7.7)\\
   $(150, 100)$ &      &0.0 (0.0)	 &0.0 (0.0) &1.1 (1.4)  &1.2 (1.9) &49.4 (6.7) &48.1 (5.1)  &50.2 (6.4) &49.4 (6.1)\\

   & & \multicolumn{8}{c}{\textbf{Scenario 3}}\\

   $(100, 50)$ &        &0.0 (0.0)	 &0.7 (1.0) &14.6 (4.6)  &1.4 (1.8) &50.2 (6.4) &50.5 (6.6)  &49.9 (6.2) &50.4 (5.8)\\
   $(150, 50)$ &        &0.0 (0.2)	 &1.0 (1.5) &17.6 (5.9)  &1.3 (1.8) &50.9 (6.5) &50.4 (5.9)  &49.9 (6.1) &49.6 (5.7) \\
   $(100, 100)$ &      &0.0 (0.0)	 &0.0 (0.3) &7.6 (3.3)    &0.9 (1.7) &50.8 (5.7) &49.8 (6.0)  &50.5 (6.1) &49.6 (7.6)\\
   $(150, 100)$ &      &0.0 (0.0)	 &0.0 (0.3) &10.2 (4.9)  &1.1 (1.7) &49.0 (5.9) &50.5 (5.9)  &49.6 (6.5) &48.4 (9.8)\\

   & & \multicolumn{8}{c}{\textbf{Scenario 4}}\\

   $(100, 50)$ &        &0.0 (0.2)	 &0.2 (0.5) &2.9 (2.5)  &1.6 (2.1) &50.7 (5.9) &50.0 (5.4)  &48.9 (6.1) &50.6 (6.8)\\
   $(150, 50)$ &        &0.0 (0.0)	 &0.1 (0.3) &1.4 (1.6)  &1.5 (2.3) &50.1 (6.2) &50.0 (6.1)  &49.7 (6.6) &50.1 (5.7)\\
   $(100, 100)$ &      &0.0 (0.0)	 &0.0 (0.0) &0.5 (0.9)  &1.2 (1.8) &49.6 (7.1) &49.7 (5.8)  &49.9 (6.9) &48.3 (9.1)\\
   $(150, 100)$ &      &0.0 (0.0)	 &0.0 (0.0) &0.1 (0.3)  &1.1 (1.4) &49.9 (6.5) &49.6 (5.5)  &50.2 (5.4) &49.5 (6.9)\\
\bottomrule[2pt]
\end{tabular*}}
\end{table}

Table \ref{tab:Classification} shows the misclassification  rates  averaged over  100 replications  for Scenario 1 - Scenario 4 with $n_1=n_2=30$, from which we can clearly see that SDNCMV performs slightly better than RF, SVM and PLR while shows overwhelming superiority over the vector-valued competitors (vRF, vSVM, vPLR) and MLDA in various scenarios. The methods MLDA, vSVM, vPLR and vPLR yield results close to a coin toss. The proposed  SDNCMV, in contrast, has misclassification rates 0.0\% in all simulation settings. RF, SVM and PLR also perform well, and RF seems perform better than SVM and PLR. This indicates that the constructed network strength variables are powerful for classification of imaging-type data.

With  fixed $p$, increasing $q$ from 50 to 100 results in much better performance for SDNCMV, RF, SVM and PLR  while with fixed $q$, increasing $p$ from 100 to 150 results in worse performance. {A question naturally aries, why the MLDA, vSVM, vPLR, vRF perform so poorly? In the data generating procedure, note that the data in two groups both have mean zero, and MLDA, vSVM, vPLR, vRF rely on the mean difference for classification. Our proposal provides an effective solution to a rarely studied problem: when two multivariate populations share the same or similar  mean but different covariance structure, with the observed training data, how to classify the test samples to the correct class? Most existing literature on classification assume that the two populations have mean difference but common covariance matrix.}

\subsection{Network Comparison Assessment}\label{sec:SS-3}

In order to assess the performance of differential network estimation, we compare our method with two joint multiple matrix Gaussian graphs estimation approaches proposed by \cite{Zhu2018multiple}, which are denoted as Non-convex and Convex. \cite{Zhu2018multiple} compared the performance of differential network estimation between their methods and some state-of-the-art  vector-valued methods. They concluded that the former is better, so we do not compare our method with any vector-valued approaches in this study.

\begin{table}[h!]\caption{{TPR, TNR and TDR averaged of the 100 replications (\%) for Scenario 1-4}}\label{tab:network}
\centering
\scalebox{0.86}{
\renewcommand{\arraystretch}{1.2}
\begin{tabular*}{18.3cm}{ccrrrrrrrrrrrrrrrrr}
\toprule[2pt]
 & &     \multicolumn{3}{c}{$(p, q)=(100, 50)$} & & \multicolumn{3}{c}{$(p, q)=(150, 50)$}  & &   \multicolumn{3}{c}{$(p, q)=(100, 100)$} & & \multicolumn{3}{c}{$(p, q)=(150, 100)$}  \\
  \cline{3-5}  \cline{7-9}  \cline{11-13} \cline{15-17}
  Methods && TPR  & TNR & TDR & & TPR   & TNR &TDR & & TPR   & TNR &TDR  & & TPR   & TNR &TDR\\
   \hline
   & & \multicolumn{15}{c}{\textbf{Scenario 1}}\\

   SDNCMV&        &90.8 	&99.9 &91.8 & &83.2 & 99.9 &90.2  & &96.1 &99.9 &97.7 & &89.9 &99.9 &96.3\\
   Non-convex&    &88.7	&95.8 &38.7 & &77.7 & 99.3 &37.4  & &99.7 &98.9 &37.3 & &98.4 &99.2 &37.3\\
   Convex&           &94.7	&98.7 &36.4 & &88.5 & 99.2 &36.0  & &99.9 &98.6 &35.5 & &99.6 &99.1 &36.1\\

   & & \multicolumn{15}{c}{\textbf{Scenario 2}}\\

   SDNCMV&        &80.7	&99.8 &91.1 & &71.0 & 99.9 &90.1  & &90.0 &99.9 &98.1 & &79.0 &99.9 &98.6\\
   Non-convex&    &88.7	&98.6 &32.2 & &64.4 & 96.6 &17.5  & &78.6 &95.3 &16.6 & &88.3 &94.9 &19.5\\
   Convex&           &87.3	&98.4 &29.9 & &64.8 & 96.6 &19.1  & &86.7 &94.8 &12.5 & &90.2 &95.3 &13.7\\

   & & \multicolumn{15}{c}{\textbf{Scenario 3}}\\

   SDNCMV&        &91.2	&99.9 &92.6 & &80.7 & 99.9 &94.5  & &97.2 &99.9 &96.5 & &89.3 &99.9 &98.5\\
   Non-convex&    &88.7	&98.6 &32.2 & &80.8 & 99.1 &32.4  & &99.3 &98.4 &32.1 & &98.6 &98.9 &32.8\\
   Convex&           &94.4	&98.0 &27.2 & &89.3 & 98.8 &27.4  & &99.1 &98.2 &29.4 & &98.2 &98.8 &30.0\\

   & & \multicolumn{15}{c}{\textbf{Scenario 4}}\\

   SDNCMV&        &80.6	&99.8 &90.4 & &80.8 & 99.7 &82.3  & &90.7 &99.9 &98.1 & &86.2 &99.9 &95.0\\
   Non-convex&    &88.7	&98.6 &32.2 & &89.9 & 90.0 &11.9  & &91.0 &94.1 &11.7 & &88.8 &94.6 &20.3\\
   Convex&           &87.3	&98.0 &29.9   & &88.5 & 92.7 &16.5  & &87.1 &94.7 &9.9 & &81.5 &95.9 &12.4\\
\bottomrule[2pt]
\end{tabular*}}
\end{table}

\begin{figure}
 \centerline{\includegraphics[width=14cm,height=17cm]{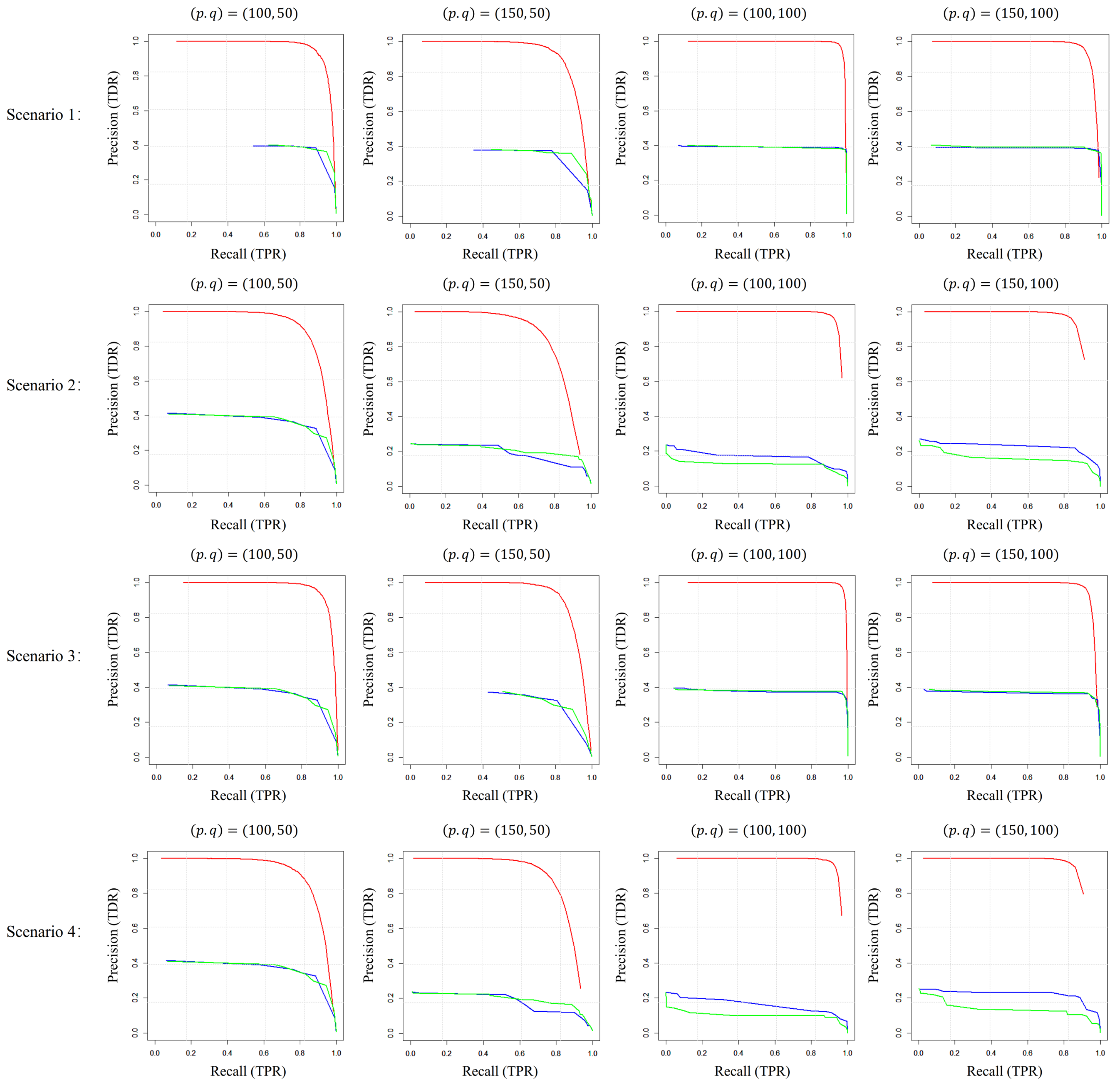}}
 \caption{Precision-Recall curve for Scenario 1-4 with varied combinations of $(p, q)$,  $n_1=n_2=30$. Red line represents SDNCMV, Blue line represents Non-convex method, and Green line represents Convex methods. }
 \label{fig:PRcurve}
\end{figure}

The criteria that we employ to evaluate the performance of  different approaches are true positive rate (TPR), true negative rate (TNR) and true discovery rate (TDR). Denote the  true differential matrix by $\bDelta=(\delta_{ij})$  and its estimate by $\hat\bDelta=(\hat\delta_{ij})$. Then, the TPR, TNR and TDR are defined as:
\[
\text{TPR}=\frac{\#\{(i,j):\delta_{ij}\ne 0\cap\hat{\delta}_{ij}\ne 0\}}{\#\{(i,j):\delta_{ij}\ne 0\}}, \ \ \
\text{TNR}=\frac{\#\{(i,j):\delta_{ij}= 0\cap\hat{\delta}_{ij}= 0\}}{\#\{(i,j):\delta_{ij}= 0\}},
\]
\[
\text{TDR}=\frac{\#\{(i,j):\delta_{ij}\ne 0\cap\hat{\delta}_{ij}\ne 0\}}{\#\{(i,j):\hat{\delta}_{ij}\ne 0\}}.
\]
 To evaluate the performance of the differential network estimators by these methods, we present the average values of TPR, TNR and TDR over 100 replications, as well as the Precision-Recall (PR) curves.

Table \ref{tab:network} shows the TPRs, TNRs and TDRs by different methods averaged over 100 replications for Scenario 1 - Scenario 4 with $n_1=n_2=30$. From Table \ref{tab:network}, it can be seen that the three methods have comparable TPRs in all scenarios, while SDNCMV has comparable TNRs with Non-convex method, which are almost all higher than those of the Convex method. The superiority of SDNCMV  is clearly shown from the TDRs, which are higher than those of the Non-convex and Convex methods by a large margin. Figure \ref{fig:PRcurve} show the Precision-Recall curves for Scenario 1-4 with varied combinations of $(p, q)$,  $n_1=n_2=30$, where red line represents SDNCMV, blue line represents Non-convex method, and green line represents Convex methods. From the PR curves, we can see the great advantages of SDNCMV over the Non-convex and Convex methods in terms of differential network structure recovery.

In summary, the simulation results show that the our method SDNCMV outperforms its competitors in terms of both classification accuracy and network comparison power,  illustrating its advantage in various scenarios. %In the supplementary materials, we present the remaining simulation results, where one would have more complete picture of the advantages of our approach.}

\section{The fMRI Data of Alzheimer's disease}

\subsection{Description of the dataset and the preprocessing procedures}
{ We applied the method to Alzheimer's Disease Neuroimaging Initiative (ADNI) study. The  ADNI  was  launched  in 2003  as  a  public-private  partnership,  led  by  Principal  Investigator    Michael    W.    Weiner,  MD.  One of the main purposes of the ADNI project is to examine differences in neuroimaging between Alzheimer's
Disease(AD) patients and normal controls
(NC). Data used in our analysis were downloaded from ADNI
website (http://www.adni.loni.usc.edu) and included resting state fMRI (rs-fMRI) images collected at screening for AD and CN participants. A T1-weighted high-resolution anatomical image
(MPRAGE) and a series of resting state functional images were
acquired with 3.0 Tesla MRI scanner (Philips Systems) during
longitudinal visits. The rs-fMRI scans were acquired with 140 volumnes, TR/TE = 3000/30 ms, flip angle of 80 and effective voxel resolution of 3.3x3.3x3.3 mm. More details can be found at ADNI website (\url{http://www.adni.loni.usc.edu}). Quality control was performed on the fMRI images both by following the Mayo clinic
quality control documentation (version 02-02-2015) and by visual
examination. After the quality control, 61 subjects were included for the analysis including $n_1=30$ AD patients and $n_2=31$ NC (normal control) subjects. For gender distribution, there are 14 (47\%) males for AD, and 14 (45\%) males for CN. The mean (SD) of age for each group is 72.88 (7.12) for AD and 74.38 (5.93) for CN).

Standard procedures were taken to preprocess the rs-fMRI data. Skull stripping was conducted on the T1 images to remove extra-cranial material. The first 4 volumes of the fMRI were removed to stabilize the signal, leaving 136 volumes for subsequent prepossessing. We registered each subject's anatomical image to the 8th volume of the slice-time-corrected functional image and then the subjects' images were normalized to MNI standard brain space. Spatial smoothing with a 6mm FWHM Gaussian kernel and motion corrections were applied to the function images. A validated confound regression approach \citep{satterthwaite2014linked,WangAn2016,kemmer2015network} was performed on each subject's  rs-fMRI time series data to remove the potential confounding factors including motion parameters, global effects, white matter (WM) and cerebrospinal fluid (CSF) signals. Furthermore, motion-related spike regressors were included to bound the observed displacement and the functional time series data were band-pass filtered to retain frequencies between 0.01 and 0.1 Hz which is the relevant range for rs-fMRI.

To construct brain networks, we considered Power's 264 node system \citep{power2011functional} that provides good coverage of
the whole brain. Therefore, the number of brain regions is  $p = 264$, and the temporal dimension is  $q = 136$. For each subject, we first an efficient algorithm implemented by \textsf{R} package ``DensParcorr" \citep{WangAn2016} to obtain the partial correlations between each pair of the 264 brain regions.

}

\subsection{Classification results of the fMRI data}

{ We applied the proposed SDNCMV method and other methods to classify AD and CN group based on the rs-fMRI connectivity.
%We observe that the measured value of some brain regions in certain subjects remain unchanged in the temporal dimension. We omit these brain regions at these time points and then obtain the partial correlations between each pair of the remaining brain regions. \textbf{We further deem that the partial correlations between the initially omitted brain nodes and the other brain regions are all zero???.} This involves 2 brain regions of one subject in AD group and 4 brain regions of  four subjects in NC group.
Since the number of  between-region network
measures is very large, i.e., $p\times(p-1)/2=34716$, we first adopt a Sure Independence Screening procedure \citep{Fan2008Sure} on the connectivity variables to avoid possible overfitting and improve computational efficiency.} We used the \textsf{R} package ``SIS" to filter out 85\% of the  variables, and retained the remaining  $5207$ variables in the active set. For a fair comparison, we adopt the same  $5207$ variables  for the methods SDNCMV, RF, SVM, PLR. For the vector-valued methods vRF, vSVM and vPLR, we applied the screening methods to filter 85\% of the total $p*q=35904$. The MLDA is computationally intractable for this real dataset and is thus omitted.  For all the methods, we take  $4$ confounders into account:  age, education level, gender, and APOlipoprotein E (APOE).

\begin{figure}[H]
	\centerline{\includegraphics[width=12cm,height=6cm]{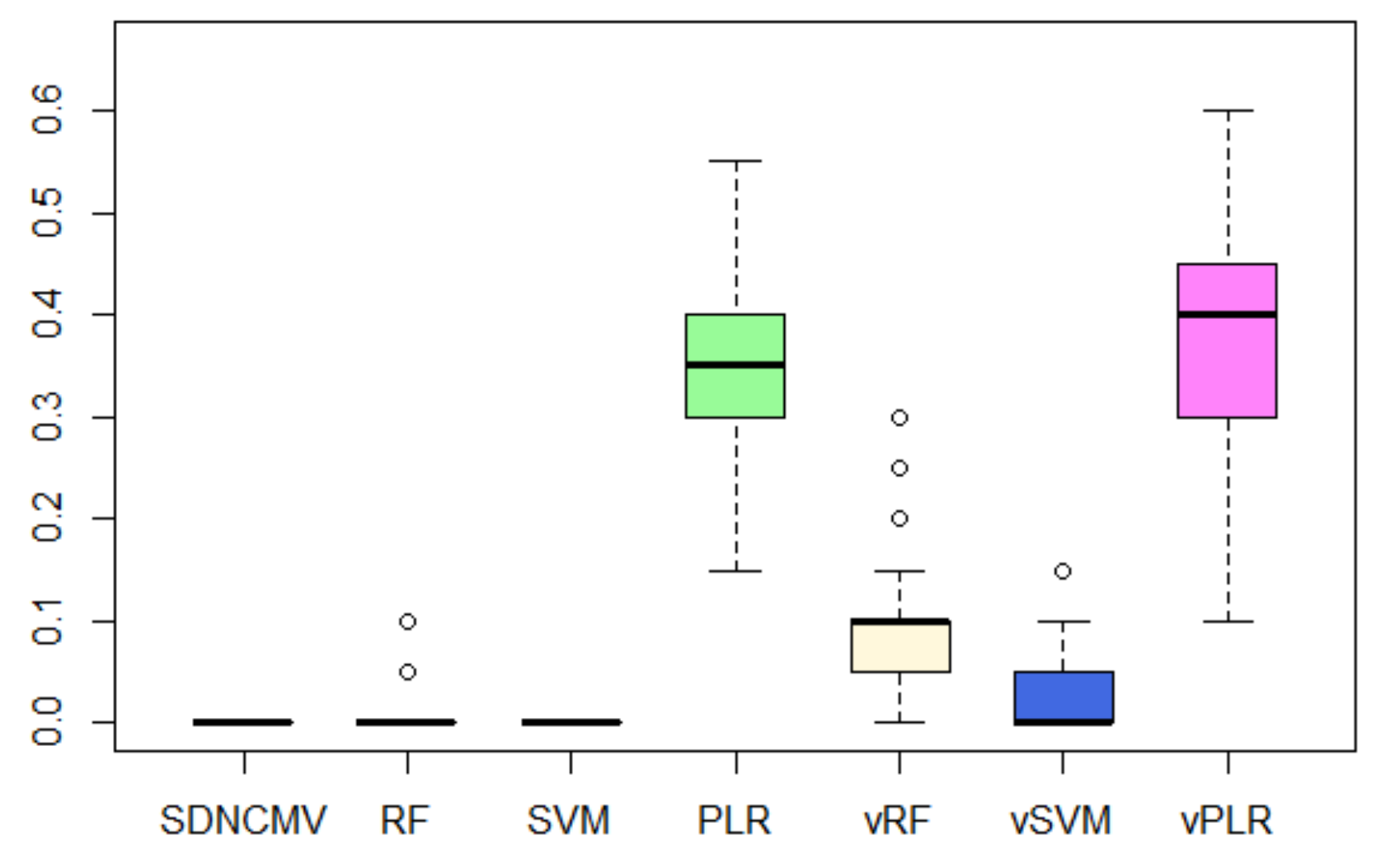}}
	\caption{Boxplot of classification errors for the fMRI Data of Alzheimer's disease over 100 bootstrap replicates.
	}
	\label{fig:realdataclass}
\end{figure}

To assess the classification accuracy of different methods, we randomly select 20 subjects from the AD group and 21 subjects from the NC group as training subjects, leaving 10 subjects in each group as the test subjects. That is, we had 41 subjects in the training set and 20 in the test set. We repeat this random sampling procedure for 100 times and report the average misclassification errors in Table \ref{tab:realdata}, and further show the boxplots of the errors in Figure \ref{fig:realdataclass}.  From Table \ref{tab:realdata}, we can see that the proposed SDNCMV does not make any mistake in classifying the test subjects, neither does the SVM method. RF method also performs very well, compared with the  vector-valued methods. In summary, the proposed SDNCMV has comparable performance with these well accepted machine learning methods.

\begin{table}[!h]\caption{{Averaged misclassification rates \% (standard errors) over 100 bootstrap replicates for the { ADNI} fMRI Data}}\label{tab:realdata}
\centering
\scalebox{1}{
\renewcommand{\arraystretch}{1.2}
\begin{tabular}{p{1.2cm}<{\centering}p{2cm}<{\centering}p{1.5cm}<{\centering}p{1.5cm}<{\centering}p{1.5cm}<{\centering}p{1.5cm}<{\centering}p{1.5cm}<{\centering}p{1.8cm}<{\centering}}
\toprule[2pt]

   Method  & SDNCMV   & RF  &SVM & PLR  & vRF  &vSVM & vPLR  \\
   Error   & 0.0 (0.0) &1.3 (2.7) &0.0 (0.0)  &34.8 (9.0) &8.1 (6.5) &2.6 (3.5) &37.4 (10.3)  \\

\bottomrule[2pt]
\end{tabular}}
\end{table}

\subsection{Network comparison results of the fMRI data}

\begin{figure}
	\centerline{\includegraphics[width=11.4cm,height=6.8cm]{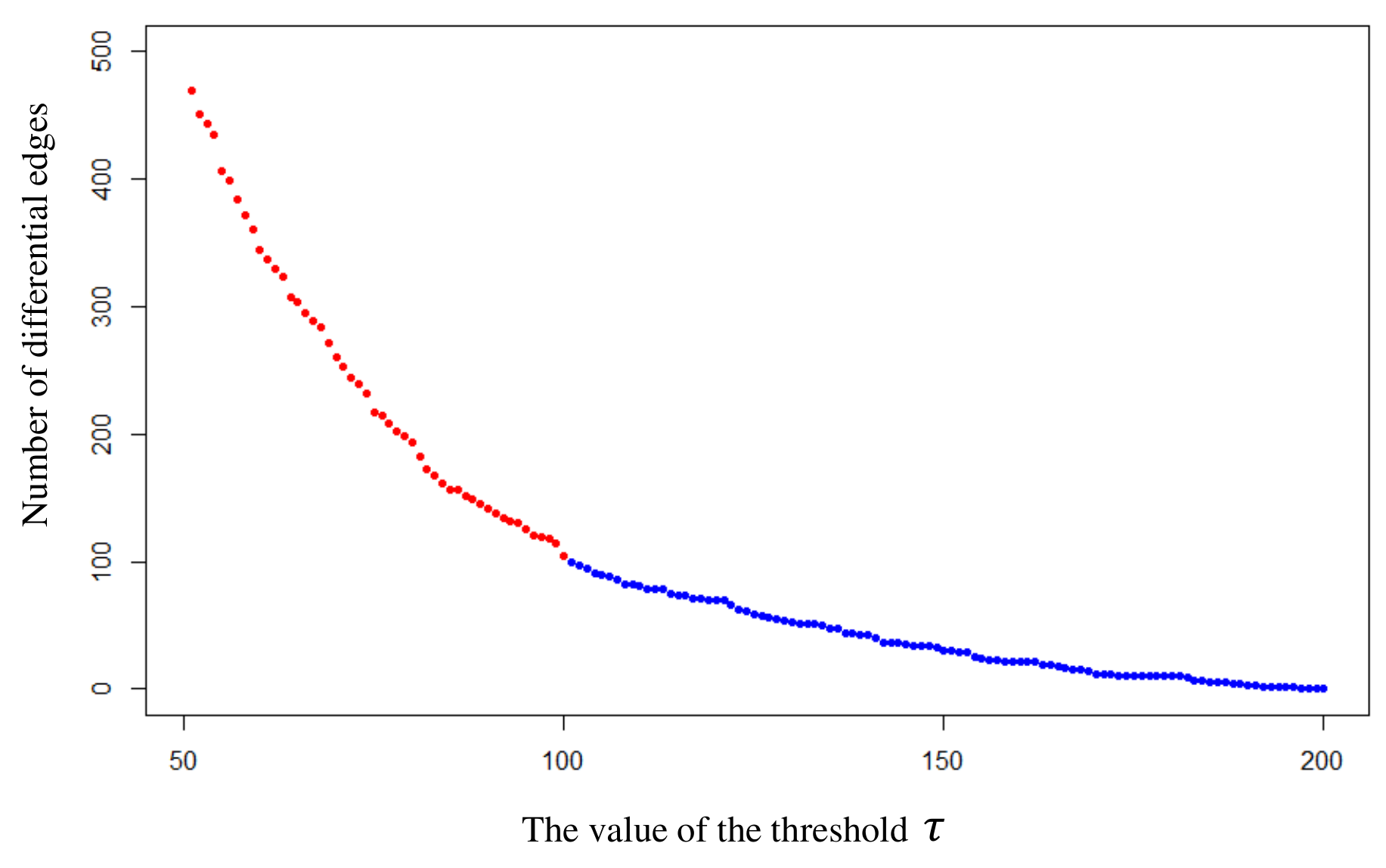}}
	\caption{Scree plot for the method SDNCMV}
	\label{fig:scatter}
\end{figure}

\begin{figure}
	\centerline{\includegraphics[width=15.6cm,height=6cm]{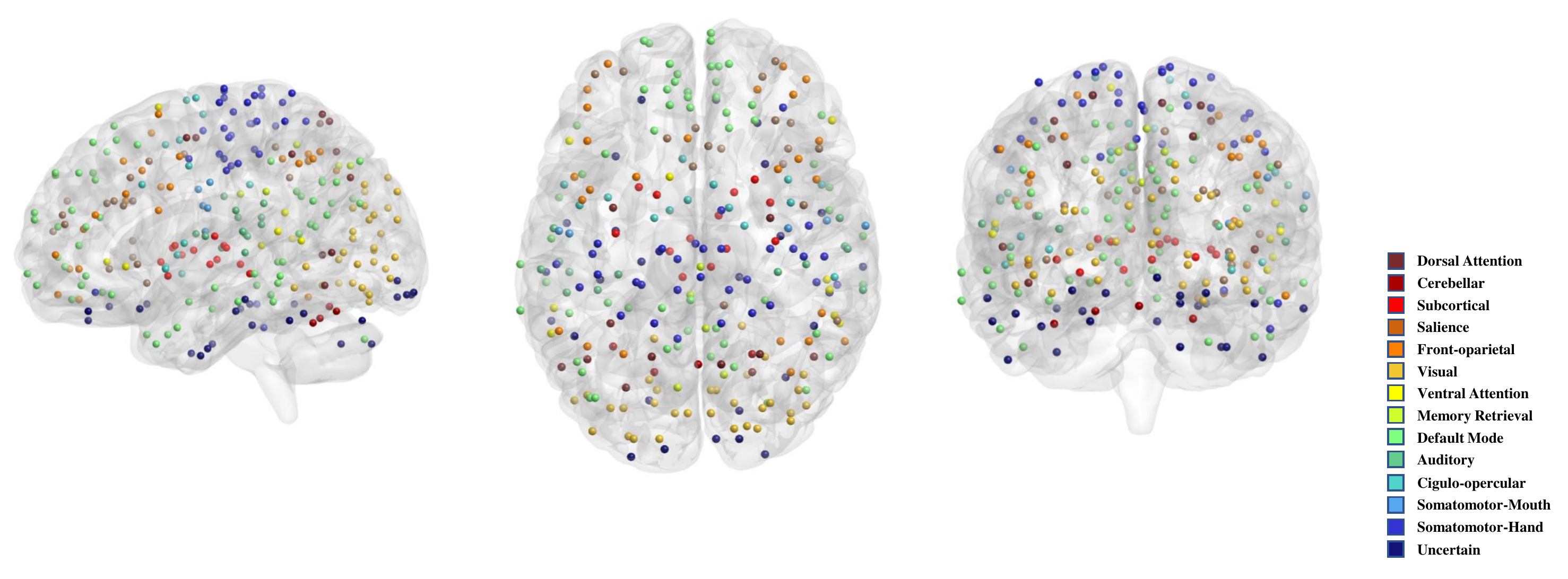}}
	\caption{Functional modules for the 264 nodes}
	\label{fig:modules}
\end{figure}

\begin{table}[h!]\caption{{The differential edges identified by SDNCMV whose number of occurrences over $B=200$ bootstrap replicates are in top 15. }}\label{tab:4}
\centering
\scalebox{1}{
\renewcommand{\arraystretch}{1.2}
\begin{tabular}{p{10cm}p{1.5cm}<{\centering}}
\toprule[2pt]

   Differential Edges                                                                             &   Occurrences   \\ \hline
   Somatomotor-Hand $\leftrightarrow$ Ventral Attention   &198  \\
   Cingulo-opercular $\leftrightarrow$ Uncertain       &197 \\
   Default Mode $\leftrightarrow$ Dorsal Attention				&192 \\
   Somatomotor-Mouth $\leftrightarrow$ Cingulo-opercular     &190 \\
   Default Mode $\leftrightarrow$ Default Mode                                &188 \\
   Default Mode $\leftrightarrow$ Dorsal Attention                          &186 \\
   Somatomotor-Hand $\leftrightarrow$ Visual                  &185 \\
   Somatomotor-Hand $\leftrightarrow$ Default Mode    &183 \\
   Front-oparietal $\leftrightarrow$ Subcortical           &183 \\
   Default Mode $\leftrightarrow$ Dorsal Attention   		      &182 \\
   Cingulo-opercular $\leftrightarrow$ Subcortical      &175 \\
   Default Mode $\leftrightarrow$ Salience				      &173 \\
   Somatomotor-Hand $\leftrightarrow$ Default Mode    &170 \\
   Default Mode $\leftrightarrow$ Visual 					      &170 \\
   Uncertain $\leftrightarrow$ Salience 					     &169 \\

\bottomrule[2pt]
\end{tabular}}
\end{table}

\begin{figure}
	\centerline{\includegraphics[width=15.6cm,height=13.6cm]{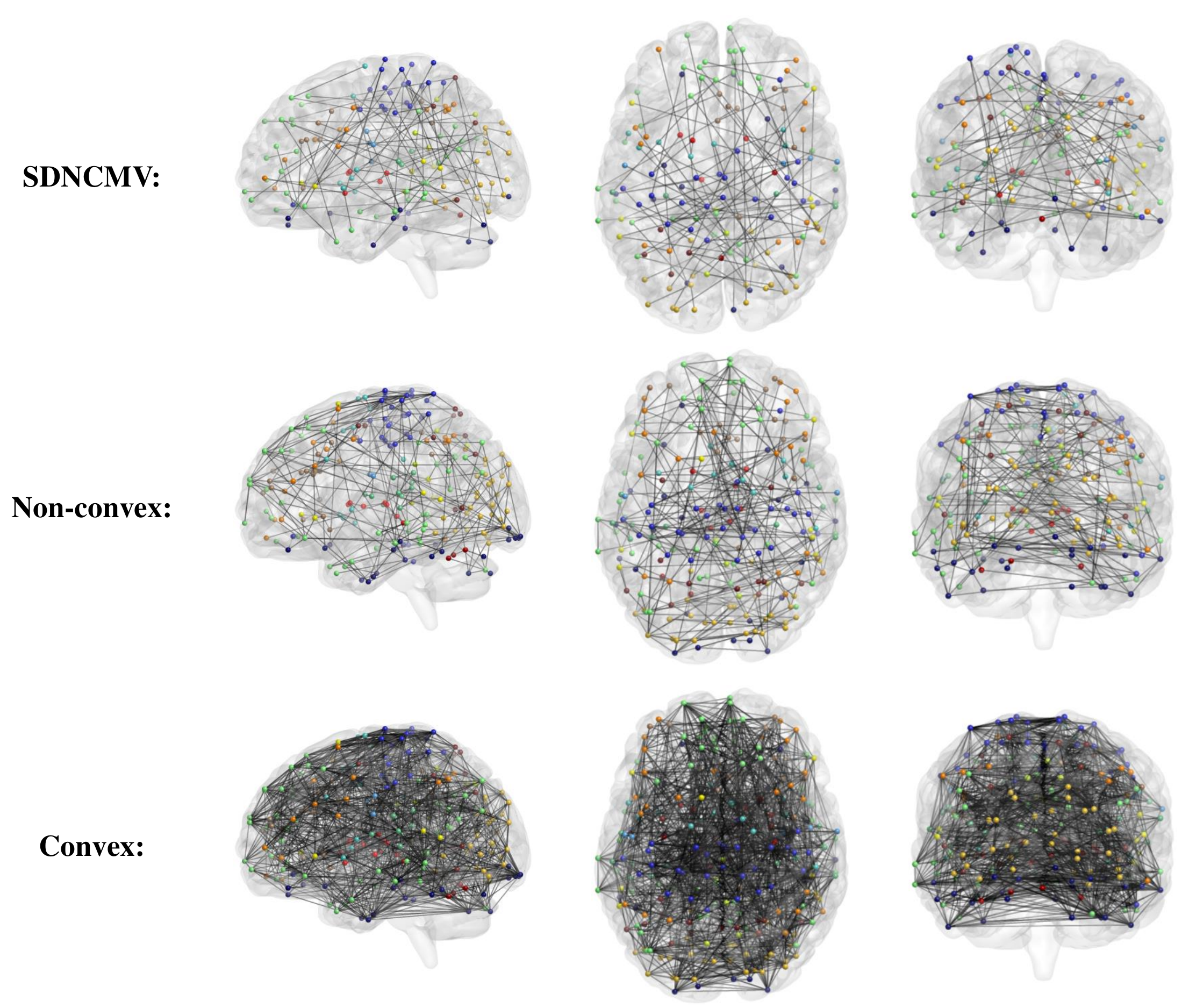}}
	\caption{Differential edges identified by various methods for the AD resting-state fMRI data. }
	\label{fig:brainnet}
\end{figure}

For network comparisons  between the AD and NC group, we used all the 30 AD subjects and 31 NC groups. We assigned the 264 nodes in the Power's sytem to 14 functional modules. Figure \ref{fig:modules} visualized the location and number of nodes for each functional module and nodes with different colors indicated that they belonged to different function modules. For more information, see \url{https://github.com/brainspaces/power264}. All the brain visualizations in this article were created using BrainNet Viewer proposed by \cite{Xia2013BrainNet}.  For the SDNCMV method, we adopt the same screening procedure as described in the last section and we determine the threshold $\tau$ as follows. We draw a scree plot in Figure \ref{fig:scatter}, the horizontal axis is the value of $\tau$ and the vertical axis is the number of the estimated differential edges by SDNCMV. From  Figure \ref{fig:scatter} , we can see that when $\tau$ is greater than about 100, the change in the number of differential edges began to be stable. Therefore, we selected 100 as the predetermined value for  threshold  $\tau$. For the Non-convex and Convex methods, the tuning parameter $\lambda$ was selected by minimizing a prediction criterion by using five fold cross-validation as described in \cite{Zhu2018multiple}.  Finally, there were totally 105 differential edges identified by SDNCMV, 3316 differential edges identified by Non-convex and 14826 differential edges identified by Convex. The convex and non-convex methods select so many differential edges
that we only show the top 10\% edges with the largest absolute values in Figure \ref{fig:brainnet}. As we can see from Figure \ref{fig:brainnet}, the differential edges identified by the Non-convex and Convex methods were too dense to  be {biologically interpretable. When examining the 105 differential edges identified by the proposed SDNCMV, we find that they mainly involve nodes located in the Somatomotor-Hand, Default Mode, and Cingulo-opercular modules.} In Table \ref{tab:4}, we showed the top 15 differential edges identified by SDNCMV and their number of occurrences over 200 bootstrap replicates. The Somatomotor-Hand is { the hand movement controlling part }of the somatic motor area which occupies most of the central anterior gyrus and executes movements selected and planned by other areas of the brain. \cite{Suva1999motor} indicated that the Somatomotor area significantly affected AD and suggested that motor dysfunction occurs in late and terminal stages of AD. The default mode is a group of brain regions that show lower levels of activity when we are engaged in a particular task like paying attention, but higher levels of activity when we are awake and not involved in any specific mental exercise. Abundance of literature has shown that default mode changes are closely related to AD \citep{Grieder2018default, Banks2018, Pei2018}. Cingulo-opercular are composed of anterior insula/operculum, dorsal anterior cingulate cortex, and thalamus, and the function of Cingulo-opercular has been particularly difficult to characterize due to the network's pervasive activity and frequent co-activation with other control-related networks. Nevertheless, some scholars have studied its relationship with Alzheimer's disease. For example, \cite{Shankar2020} found that loss of segregation in Cingulo-opercular network was associated with apathy in AD and suggested that network-level changes in AD patients may underlie specific neuropsychiatric symptoms. In summary, the findings from the proposed SDNCMV are consistent with evidences from a wide range of neuroscience and clinical studies.

\section{Summary and Discussions}

In the article, we focus on the network comparison and two-class classification problem for matrix-valued fMRI data. We propose an effective method which fulfills the goals of network comparison and classification simultaneously. Numerical study shows that the SDNCMV performs advantageously than the state-of-the-art methods for classification and network comparison methods. We also illustrate the utility of the  SDNCMV  by analyzing the fMRI Data of Alzheimer's disease. Our method can be widely applied in brain connectivity alteration detection and image-guided medical diagnoses of complex diseases.

The matrix normal assumption of the matrix-valued data can be relaxed so that we only assume a  Kronecker product covariance structure. In the case that outliers or heavy-tailedness of data exist, we can propose a similar robust version of  SDNCMV. In detail, the sample covariance matrix in (\ref{equ:samplecov}) can be replaced by some robust estimators such as the adaptive Huber estimator and the median of means estimator \citep{avella2018robust}. Another way to relax the matrix-normal assumption is the matrix-non-paranormal distribution in \cite{NingHigh2013},  which can be viewed as a latent-variable model. That is, the latent variables $f(\Xb)$ follow a matrix-normal distribution and must be symmetric, while the observed variables $\Xb$ need not be symmetric. In this case, we adopt the Kendall's tau based correlation estimators in (\ref{equ:samplecov}), which has been widely studied in literature such as \cite{liu2012high,He2018Variable,fan2018,He2019Robust,Yu2019Robust}, etc.

In the second-stage model, we focus on a binary outcome response and adopt the logistic regression for classification. It is straightforward to extend to a general  clinical outcome response with a more flexible Generalized Linear Model (GLM).

Some future work remains to develop inferential tools on the significance of a selected edge, e.g., in both global testing and multiple testing problems based on the logistic regression model in (\ref{equ:logit}).

\section*{Acknowledgements}
This work was supported by grants from the National Natural Science
Foundation of China [grant number  11801316, 81803336, 11971116]; Natural Science Foundation of Shandong Province [grant number ZR2018BH033, ZR2019QA002], and National Center for Advancing Translational Sciences (grant number UL1TR002345 to L.L. { Data collection and   sharing for  this   project   was  funded   by  the   Alzheimer's   Disease Neuroimaging  Initiative  (ADNI)  (National  Institutes  of  Health  Grant  U01  AG024904)  and DOD  ADNI  (Department  of  Defense  award  number  W81XWH-12-2-0012).  ADNI  is  funded by  the   National  Institute  on  Aging,  the  National  Institute  of  Biomedical  Imaging  and Bioengineering, and through generous contributions from the following: AbbVie, Alzheimer's  Association;  Alzheimer's  Drug  Discovery  Foundation;  Araclon  Biotech;  BioClinica,  Inc.; Biogen;   Bristol-Myers   Squibb   Company;   CereSpir,   Inc.;   Cogstate;   Eisai   Inc.;   ElanPharmaceuticals,  Inc.;  Eli  Lilly  and  Company;  EuroImmun;  F.  Hoffmann-La  Roche  Ltd and its  affiliated  company  Genentech,  Inc.;  Fujirebio;  GE  Healthcare;  IXICO  Ltd.;  Janssen Alzheimer    Immunotherapy    Research    $\&$    Development,    LLC.;    Johnson      $\&$    Johnson Pharmaceutical  Research    $\&$  Development  LLC.;  Lumosity;  Lundbeck;  Merck    $\&$  Co.,  Inc.; Meso  Scale  Diagnostics,  LLC.;  NeuroRx  Research;  Neurotrack  Technologies;  Novartis Pharmaceuticals Corporation; Pfizer Inc.; Piramal Imaging; Servier; Takeda Pharmaceutical Company;  andTransition  Therapeutics.  The  Canadian  Institutes  of  Health  Research  is providing  funds  to  support  ADNI  clinical  sites  in  Canada.  Private  sector  contributions  are facilitated by the Foundation for the National Institutes of Health (www.fnih.org). The grantee organization is the Northern California Institute for Research and Education, and the study is coordinated by the Alzheimer's Therapeutic Research Institute at the University of Southern California.  ADNI  data  are  disseminated   by  the   Laboratory  for  NeuroImaging  at  the University of SouthernCalifornia.}).

\bibliographystyle{model2-names}
\bibliography{ref}
	
\end{document}